# Cosmological parameters from angular correlations between QSOs and galaxies

**Matthias Bartelmann**

Max-Planck-Institut für Astrophysik, Postfach 1523, D–85740 Garching, FRG

28 July 1994

**Abstract.** The angular two-point correlation function between background QSOs and foreground galaxies induced by gravitational lensing is derived. It is shown that the shape of this correlation function depends sensitively on the spectrum of the density fluctuations in the Universe, thus providing a possibility to distinguish between different models for the spectrum. Using numerical large-scale structure simulations, I estimate that the QSO-galaxy correlation function can be measured from galaxy counts down to $\simeq 21^{\mathrm{st}} \ldots 22^{\mathrm{nd}}$ magnitude in fields with radius $\lesssim 25'$ around $50 \ldots 100$ QSOs with redshift $z \geq 1$. Since the QSO-galaxy correlation function is proportional to $(-a-1)b$, where $b$ is the biasing factor of galaxy formation and $a$ is the effective slope of the QSO number counts, steep number counts are favorable for this kind of analysis. I show that $a = -4 \ldots -5$ can be achieved with the 1-Jansky sample of radio-loud QSOs when the double-waveband magnification bias is employed. Moreover, the cross-correlation analysis allows to determine the galaxy-formation bias factor $b$.

## 1 Introduction

It was shown in a sequence of previous papers that there is evidence for highly significant correlations on angular scales of several 10 arc minutes between radio-loud, high-redshift QSOs and foreground optical galaxies (Fugmann 1992, Bartelmann & Schneider 1993b), IRAS galaxies (Bartelmann & Schneider 1994a) and diffuse X-ray emission (Bartelmann, Schneider & Hasinger 1994b). In an earlier study, it was demonstrated on the basis of numerical large-scale structure models that gravitational lensing by large-scale structures provides a viable explanation for such correlations (Bartelmann & Schneider 1993a). The basic underlying idea was the following. Dark matter, which is usually assumed to dominate the mass density in the Universe, must be inhomogeneously distributed in space, because otherwise structure formation can hardly be understood. A matter distribution with spatially varying density acts as a gravitational lens, magnifying and distorting images of background sources. In a flux-limited sample of background sources such as high-redshift QSOs, such magnified sources will preferentially be included. This effect,

---

*Send offprint requests to:* M. Bartelmann

termed magnification bias, is extensively described in the literature (see, amongst numerous others, Schneider, Ehlers, & Falco 1992, Chap.12, Narayan & Wallington 1993, and references therein). This implies that sources from a flux-limited sample will preferentially be found in the background of matter overdensities. From the biasing hypothesis of galaxy formation (e.g., Kaiser 1984, Dekel & Rees 1987), it is expected that galaxies form preferentially where the dark matter is overdense. Combined with the magnification bias caused by these overdensities to a flux-limited sample, this implies that such background sources are preferentially observed where the foreground galaxy density is enhanced, and this yields gravitational-lensing induced correlations between background QSOs and foreground galaxies.

The occurrence of such correlations has been tested and established on a significance level of up to 99.8% in the previously cited papers. There, Spearman's rank-order correlation test was employed. This test has the advantage of being robust, parameter-free, and highly sensitive (e.g., Kendall & Stuart 1973), but it has the disadvantage that it yields a 'binary' result, since it merely answers the question of whether QSOs and galaxies can be considered correlated or not. Nevertheless, the correlation tests performed unambiguously demonstrate that such a correlation exists, although they do not allow to quantify the strength of the correlation; note that a *significant* correlation does not necessarily imply a *strong* correlation if a highly sensitive correlation test is used.

From the arguments stated above, it is clear that, if gravitational lensing is indeed the reason for the detected correlations, these prove the biasing hypothesis of galaxy formation. If this hypothesis were not at least qualitatively correct, galaxies were not correlated with dark-matter overdensities, and then the distribution of dark lenses would be random compared to the distribution of galaxies.

This paper addresses the question of how the biasing hypothesis of galaxy formation might be quantified on the basis of the gravitational-lensing interpretation of the correlations between background sources and foreground galaxies. Such a quantification is interesting in various respects. First, it constitutes a further observational test for such cosmogonic scenarios which are fairly specific in the amount of biasing they require to describe the observed galaxy distribution from the theoretically modeled dark-matter distribution (e.g., White et al. 1987, Cen & Ostriker 1992, Ostriker 1993). Second, it provides a lower limit for the mass-to-light ratio on cosmological scales beyond the cluster scale. This is because the bias factor quantifies the ratio of the fluctuations of the galaxy number density relative to the density fluctuations. Homogeneously distributed dark matter does not contribute to the magnification bias due to gravitational lensing because it does not magnify background sources, and thus only deviations of the cosmic density from its mean contribute to the effect.

In Sect.2 of the present paper, I derive the angular cross-correlation function $\xi_{\rm QG}(\phi)$ between background sources and foreground galaxies as expected from gravitational lensing by large-scale structures. This is done in close analogy to Kaiser (1992). I show that $\xi_{\rm QG}(\phi)$ is proportional to $(-a-1)$, where $a$ is the effective double-logarithmic slope of the luminosity function of the background sources, and to the effective biasing factor $b$ of the foreground galaxies on the scale considered. The shape of $\xi_{\rm QG}(\phi)$ depends strongly on the shape of the dark-matter perturbation spectrum $P_\delta(k)$. A series expansion of $\xi_{\rm QG}(\phi)$ for $\phi \ll 1$ shows how parameters determining the shape of $P_\delta(k)$ can be derived from $\xi_{\rm QG}(\phi)$. In Sect.3, I compare the analytic results obtained in the previous section with a correlation function derived from numerical simulations incorporating a simple recipe for galaxy formation. It is argued there that in reality the biasing parameter may be a function of scale, as it is in the numerical simulation. This scale-dependence of $b$ slightly steepens the correlation function. The numerical simulation is intended to demonstrate the feasibility of determining $\xi_{\mathrm{QG}}(\phi)$ from real data if sufficiently deep ($\gtrsim 21^{\mathrm{th}}$ magnitude) galaxy observations in sufficiently large ($\gtrsim 25'$) fields around $50\ldots 100$ background QSOs are available. Bootstrap error estimates are given. In Sect.4 I argue that the multiple-waveband magnification bias first discussed by Borgeest et al. (1991) effectively steepens the luminosity function to cumulative slopes of $a = -4\ldots -5$ in the 1-Jansky sample, dependent on the optical flux limit imposed. A slope as steep as that was required in earlier theoretical work (Bartelmann & Schneider 1992), and it is shown that it can indeed be achieved with real QSO samples. Sect.4 summarizes the results.

## 2 The angular QSO-galaxy correlation function from weak lensing

### 2.1 Light propagation

It was shown by Kaiser (1992) that the (two-dimensional) deflection angle of a light ray propagating through an Einstein-de Sitter background universe which is slightly perturbed by density inhomogeneities with Newtonian gravitational potential $\Phi$ is given by

$$\alpha_j(\vec{\theta}, w) = -\frac{2}{w}\int_0^w dw'(w-w')\Phi_{,j}(w'\theta_1, w'\theta_2, w'; 1-w') \ . \tag{2.1}$$

To render this paper self-contained, I re-derive this equation in Appendix A. The index on $\Phi$, preceded by a comma, denotes partial differentiation with respect to the coordinate $w_j$. $w$ is the comoving distance in units of twice the Hubble length, $\vec{\theta}$ is the angular position vector of the light ray on the observer's sky, and the argument $1-w$ of $\Phi$ is the conformal cosmological time at which the potential perturbation $\Phi$ is considered. The important approximation which entered the derivation of Eq.(2.1) was that the potential gradient is integrated along the *unperturbed* path of the light ray.

The deformation of an infinitesimal light bundle is determined by the shear tensor

$$\gamma_{jk}(\vec{\theta}, w) \equiv \frac{\partial \alpha_j(\vec{\theta}, w)}{\partial \theta_k} = -\frac{2}{w}\int_0^w dw' w'(w-w')\Phi_{,jk}(w'\theta_1, w'\theta_2, w'; 1-w') \ . \tag{2.2}$$

Let now $\mu$ be the magnification along the light ray, i.e., the solid-angle distortion of a light bundle. Then, $\langle\mu\rangle = 1$ when averaged over the whole sky. The magnification $\mu$ is determined by the determinant of $\gamma_{jk}$,

$$\mu = \frac{1}{\det(\delta_{jk} + \gamma_{jk})} \simeq 1 - \mathrm{tr}(\gamma_{jk}) \ , \tag{2.3}$$

where it was assumed in the last step that $|\gamma_{jk}| \ll 1$. The deviation of $\mu$ from unity, $\delta\mu \equiv \mu - \langle\mu\rangle = \mu - 1$, is then determined by the trace of the shear tensor, or

$$\delta\mu(\vec{\theta}, w) = -\delta_{jk}\gamma_{jk}(\vec{\theta}, w) = \frac{12}{w}\int_0^w dw' w'(w-w')\frac{\delta(w'\theta_1, w'\theta_2, w'; 1-w')}{(1-w')^2} \ , \tag{2.4}$$

where we have used Poisson's equation (A11). Eq.(2.4) relates the magnification of the light ray to the density contrast along the light ray.

## 2.2 The angular QSO-galaxy correlation function

Assume now that the observer sees a QSO population above flux threshold $S$ characterized by its number density on the sphere $n_\mathrm{Q}(S;\vec{\theta})$, and similarly a galaxy population characterized by $n_\mathrm{G}(\vec{\theta})$. Their angular cross-correlation function is

$$\xi_\mathrm{QG}(\phi) = \frac{1}{\langle n_\mathrm{G}\rangle \langle n_\mathrm{Q}(S)\rangle} \left\langle \left[n_\mathrm{Q}(S;\vec{\theta}) - \langle n_\mathrm{Q}(S)\rangle\right] \left[n_\mathrm{G}(\vec{\theta}+\vec{\phi}) - \langle n_\mathrm{G}\rangle\right]\right\rangle , \qquad (2.5)$$

where the average extends over $\vec{\theta}$, and the result does not depend on the direction of $\vec{\phi}$ due to the statistical isotropy of the QSO and galaxy distributions on the sky.

The QSO population seen by the observer depends on the unlensed QSO number counts and, since the QSO sample is flux limited, also on the magnification of the QSOs via the magnification bias. Let the unlensed cumulative number density be

$$n'_\mathrm{Q}(S') \propto (S')^a , \qquad (2.6)$$

where $S'$ is the QSO flux in absence of gravitational magnification. The power-law index $a$ is assumed to be independent of the QSO redshift, which appears to be an uncritical approximation. Due to gravitational lensing, the flux is magnified by a factor $\mu$. Therefore, $S = \mu S'$, and the observed number counts are related to the unlensed counts by

$$n_\mathrm{Q}(S) = \frac{1}{\mu} n'_\mathrm{Q}\left(\frac{S}{\mu}\right) . \qquad (2.7)$$

The factor $1/\mu$ out front accounts for the solid-angle distortion which causes the magnification; the solid angle of a magnified patch of the sky is reduced. With the power-law assumption (2.6), we then have

$$n_\mathrm{Q}(S) \propto \mu^{-a-1} S^a . \qquad (2.8)$$

If $a = -1$, the unlensed number counts are unchanged. Since the average number density of QSOs equals the unlensed one because $\langle\mu\rangle = 1$, we obtain

$$\frac{n_\mathrm{Q}(S;\vec{\theta}) - \langle n_\mathrm{Q}(S)\rangle}{\langle n_\mathrm{Q}(S)\rangle} = \mu^{-a-1} - 1 . \qquad (2.9)$$

With $\mu = 1 + \delta\mu$ and $\delta\mu \ll 1$ as expected from weak lensing (see, e.g., Jaroszyński et al. 1990, Bartelmann & Schneider 1991),

$$\mu^{-a-1} - 1 \simeq (-a-1)\delta\mu . \qquad (2.10)$$

We now assume that the galaxy number density is related to the density contrast $\delta$ by the bias factor $b$,

$$\frac{n_\mathrm{G} - \langle n_\mathrm{G}\rangle}{\langle n_\mathrm{G}\rangle} = b\delta . \qquad (2.11)$$

The effect of a possible scale dependence of the bias factor $b$ is addressed in the next section. Then, the correlation function $\xi_\mathrm{QG}(\phi)$ of Eq.(2.5) reduces to

$$\xi_\mathrm{QG}(\phi) = (-a-1)b \left\langle \delta\mu(\vec{\theta})\delta(\vec{\theta}+\vec{\phi})\right\rangle \equiv (-a-1)b\, \xi_{\mu\delta}(\phi) , \qquad (2.12)$$

where $\xi_{\mu\delta}$ is the angular cross-correlation function between magnification and density contrast.

## 2.3 Evaluation of $\xi_{\mu\delta}(\phi)$

**2.3.1 QSO weight function.** We assumed that the QSO number counts have a flux distribution with slope $a$ independent of their distance. Their distance distribution be characterized by a normalized function $W_Q(w)$. Then, the magnification $\delta\mu$ of Eq.(2.4) has to be weighted with $W_Q$, or

$$\delta\mu(\vec{\theta}) = \int_0^1 dw\, W_Q(w)\delta\mu(\vec{\theta},w)$$
$$= 12 \int_0^1 dw \int_0^w dw'\, \frac{w'(w-w')}{w} W_Q(w) \frac{\delta(\vec{w}',1-w')}{(1-w')^2} \,. \qquad (2.13)$$

Changing the order of integrations, we find

$$\delta\mu(\vec{\theta}) = 12 \int_0^1 dw'\, G_Q(w') \frac{\delta(\vec{w}',1-w')}{(1-w')^2} \,, \qquad (2.14)$$

where

$$G_Q(w) \equiv w \int_w^1 dw' \left(1 - \frac{w}{w'}\right) W_Q(w') \,. \qquad (2.15)$$

If all observed QSOs were located at the same distance $w_0$, we would have

$$W_Q(w) = \delta(w - w_0) \,, \qquad (2.16)$$

which would result in

$$G_Q(w) = H(w_0 - w)\, w \left(1 - \frac{w}{w_0}\right) \,, \qquad (2.17)$$

where $H(x)$ is Heaviside's step function.

Instead, we will adopt in the following the QSO weight function

$$W_Q(w) = C\, w^{-\gamma} \,, \quad w \in [w_0, 1] \,, \qquad (2.18)$$

where $w_0$ is the minimum distance of QSOs in the sample, which is related to the lower QSO cutoff redshift

$$z_Q = \frac{1}{(1-w_0)^2} - 1 \,. \qquad (2.19)$$

For $W_Q$ to be normalized, we require

$$C = \begin{cases} \frac{\gamma-1}{w_0^{1-\gamma}-1} & \text{for} \quad \gamma \neq 1 \\ -\frac{1}{\ln w_0} & \text{for} \quad \gamma = 1 \end{cases} \,. \qquad (2.20)$$

From Eq.(2.15), we obtain

$$G_Q(w) = \begin{cases} \frac{w}{1-w_0}\left(1 - \bar{w} + w \ln \bar{w}\right) & \text{for} \quad \gamma = 0 \\ \frac{w}{\ln w_0}\left(\ln \bar{w} - w + \frac{w}{\bar{w}}\right) & \text{for} \quad \gamma = 1 \\ \frac{w}{1-w_0^{1-\gamma}}\left[1 - \bar{w}^{1-\gamma} + \frac{1-\gamma}{\gamma} w\left(1 - \bar{w}^{-\gamma}\right)\right] & \text{otherwise} \end{cases} \,, \qquad (2.21)$$

where

$$\bar{w} \equiv \max(w, w_0) \,. \qquad (2.22)$$

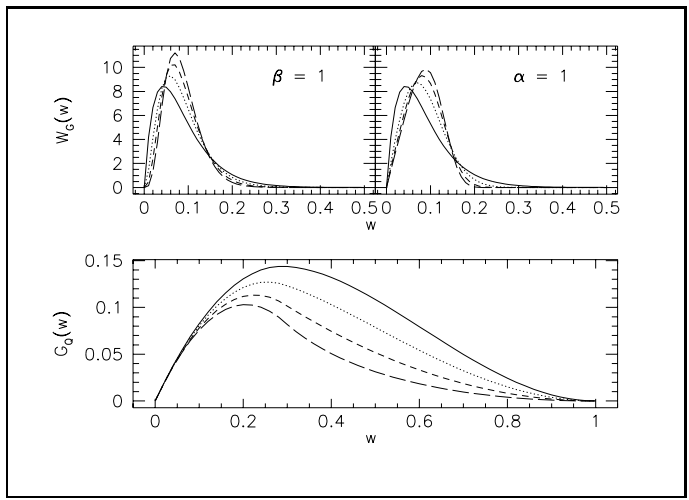

**Fig. 1.** The weight functions $W_{\mathrm{G}}(w)$ for $\beta = 1$ and $\alpha \in \{1,\ldots,4\}$ (top left frame), for $\alpha = 1$ and $\beta \in \{1,\ldots,4\}$ (top right frame), and $G_{\mathrm{Q}}(w)$ for $\gamma \in \{0,\ldots,3\}$ in the bottom frame. $z_{\mathrm{G}} = 0.2$ and $z_{\mathrm{Q}} = 1.0$ are kept fixed

**2.3.2 Galaxy weight function.** The galaxy distribution along the light ray must also be weighted by some galaxy weight function $W_{\mathrm{G}}(w)$ which is meant to mimic the distance distribution of galaxies in a magnitude-limited galaxy sample. We adopt the weight function given by Kaiser (1992),

$$W_{\mathrm{G}}(w) = \frac{\beta}{w_*\, \Gamma\left(\frac{1+\alpha}{\beta}\right)} \left(\frac{w}{w_*}\right)^\alpha \exp\left[\left(-\frac{w}{w_*}\right)^\beta\right],  \qquad (2.23)$$

which is normalized to unity. Kaiser (1992) chooses $\alpha = 1$ and $\beta = 4$. The average galaxy distance is related to $w_*$ by

$$w_{\mathrm{G}} = w_* \, \frac{\Gamma\left[(2+\alpha)/\beta\right]}{\Gamma\left[(1+\alpha)/\beta\right]}, \qquad (2.24)$$

and the average galaxy redshift $z_{\mathrm{G}}$ is determined by $w_{\mathrm{G}}$ analogous to Eq.(2.19). With Eq.(2.23), we obtain the projected density contrast

$$\delta(\vec{\theta}) = \int_0^1 dw\, \delta(\vec{w}, 1-w) W_{\mathrm{G}}(w) \,. \qquad (2.25)$$

Figure 1 shows the weight functions $W_{\mathrm{G}}(w)$ for $\beta = 1$ and $\alpha \in \{1,\ldots,4\}$ in the top left frame, for $\alpha = 1$ and $\beta \in \{1,\ldots,4\}$ in the top right frame, and $G_{\mathrm{Q}}(w)$ for $\gamma \in \{0,\ldots,3\}$ in the bottom frame. $z_{\mathrm{G}} = 0.2$ and $z_{\mathrm{Q}} = 1.0$ are kept fixed.

**2.3.3 Power spectra.** Kaiser (1992) has shown that it is most convenient to evaluate correlation functions of the type considered here in Fourier space, making use of the fact that the auto-correlation function of some quantity $p$ is the Fourier transform of the power spectrum of this quantity,

$$\xi_{pp}(\phi) = \int \frac{d^2\kappa}{(2\pi)^2} P_p(\kappa) \exp(-\mathrm{i}\vec{\kappa}\vec{\phi}) \qquad (2.26)$$

Both quantities of interest here, $\delta\mu$ and $\delta$, invoke weighted integrations over the density contrast along some light ray. The problem therefore arises of how the spectrum $P_p(\kappa)$ of a quantity $p$ which is a weighted projection of a three-dimensional quantity $f$ is related to the spectrum $P_f(k)$ of $f$. The answer is given in the appendix of Kaiser (1992), Eq.(A2). Given a statistically homogeneous and isotropic random field $f(\vec{x})$ and a weighted projection

$$p(\vec{\theta}) = \int_0^1 dw\, q(w) f(w\theta_1, w\theta_2, w) \tag{2.27}$$

of $f$, then the spectrum of $p$ is

$$P_p(\kappa) = \int_0^1 dw\, \frac{q^2(w)}{w^2} P_f\left(\frac{\kappa}{w}, 1-w\right). \tag{2.28}$$

This Fourier-space analogue of Limber's equation (e.g., Peebles 1993, Eq.(7.47)) arises if it is assumed that the $w$-integration can be decomposed into a sum over shells with width $\Delta w \ll 1$, whose contributions are mutually statistically independent such that $P_f(k) \simeq 0$ for $k \lesssim (1/\Delta w)$. For the purpose discussed here, this assumption is safely fulfilled. A typical scale in the power spectrum of the density contrast is $\simeq 50$ Mpc/$h$, which is much smaller than twice the Hubble length which corresponds to $w = 1$.

We are here after the cross-correlation function between magnification and density contrast $\xi_{\mu\delta}$. Both the magnification and the density contrast have the form of Eq.(2.27), where $12G_Q(w)/(1-w)^2$ and $W_G(w)$ adopt the role of $q(w)$, respectively. In both cases, $f(\vec{w})$ is the density contrast $\delta(\vec{w})$. Therefore, we can insert Eqs.(2.14) and (2.25) into (2.28) and obtain

$$P_{\mu\delta}(\kappa) = 12 \int_0^1 dw\, \frac{G_Q(w)W_G(w)}{w^2(1-w)^2} P_\delta\left(\frac{\kappa}{w}, 1-w\right). \tag{2.29}$$

The angular cross-correlation function $\xi_{\mu\delta}$ now follows from Eqs.(2.26) and (2.29);

$$\xi_{\mu\delta}(\phi) = 12 \int_0^1 dw\, \frac{G_Q(w)W_G(w)}{w^2(1-w)^2} \int_0^\infty \frac{\kappa d\kappa}{2\pi} P_\delta\left(\frac{\kappa}{w}, 1-w\right) \int_0^{2\pi} \frac{d\vartheta}{2\pi} \exp(-\mathrm{i}\kappa\phi\cos\vartheta), \tag{2.30}$$

Where we have introduced the angle $\vartheta$ between $\vec{\kappa}$ and $\vec{\phi}$. The rightmost integral results in the zeroth-order Bessel function of the first kind,

$$\int_0^{2\pi} \frac{d\vartheta}{2\pi} \exp(-\mathrm{i}\kappa\phi\cos\vartheta) = \mathrm{J}_0(\kappa\phi). \tag{2.31}$$

Further, we substitute $k \equiv (\kappa/w)$ and obtain

$$\xi_{\mu\delta}(\phi) = 12 \int_0^1 dw\, \frac{G_Q(w)W_G(w)}{(1-w)^2} \int_0^\infty \frac{kdk}{2\pi} P_\delta(k, 1-w) \mathrm{J}_0(wk\phi). \tag{2.32}$$

If we assume linear density evolution, we have

$$P_\delta(k, 1-w) = (1-w)^4 P_\delta(k, 0) \equiv (1-w)^4 P_\delta^{(0)}(k), \tag{2.33}$$

and thus

$$\xi_{\mu\delta}(\phi) = 12 \int_0^\infty \frac{kdk}{2\pi} P_\delta^{(0)}(k) \int_0^1 dw (1-w)^2 G_Q(w) W_G(w) \mathrm{J}_0(wk\phi). \tag{2.34}$$

We will use for CDM the linearly evolved perturbation spectrum as given by Bardeen et al. (1986). For HDM, we will instead use the spectrum as derived from the Zel'dovich approximation by Schneider & Bartelmann (1994). Since the Zel'dovich approximation affects the spectrum only for small scales (large $k$), this spectrum will also be scaled in proportion to $a^2 = (1-w)^4$ as in the linear case, but will be modified for large $k$ dependent on $1-w$.

To acquire some insight into the behaviour of $\xi_{\mu\delta}(\phi)$ for CDM and HDM spectra, we insert the model spectra

$$P^{(0)}_{\delta,\text{HDM}}(k) = A_{\text{HDM}} k \exp(-\lambda k) ;$$
$$P^{(0)}_{\delta,\text{CDM}}(k) = A_{\text{CDM}} k \frac{1}{(k^2 + k_0^2)^2} \qquad (2.35)$$

into Eq.(2.34). For the exponentially decaying spectrum, the $k$–integration is readily performed using Eq.(6.621.4) of Gradshteyn & Ryzhik (1980); it yields

$$\xi_{\mu\delta}^{\text{HDM}}(\phi) = A_{\text{HDM}} \int_0^1 dw (1-w)^2 G_Q(w) W_G(w) \frac{2\lambda^2 - (w\phi)^2}{[\lambda^2 + (w\phi)^2]^{5/2}} . \qquad (2.36)$$

For the algebraically decaying model spectrum, the $k$–integration leads to hypergeometric functions $_1F_2$; the result is therefore of little practical use. However, series expansions of the correlation functions after evaluating the integral in Eq.(2.36) for both model spectra for small $\phi$ yield

$$\xi_{\mu\delta}^{\text{HDM}}(\phi) = C_0 - C_2 \phi^2 + \mathcal{O}(\phi^3) ,$$
$$\xi_{\mu\delta}^{\text{CDM}}(\phi) = D_0 - D_1 \phi + D_2 \phi^2 + \mathcal{O}(\phi^3) , \qquad (2.37)$$

where the constants $C_i$ and $D_j$ are

$$C_0 = A_{\text{HDM}} \frac{1}{\pi \lambda^3} \int_0^1 dw\, K(w) ,$$
$$C_2 = A_{\text{HDM}} \frac{3}{\pi \lambda^5} \int_0^1 dw\, w^2 K(w) ,$$
$$D_0 = A_{\text{CDM}} \frac{1}{8 k_0} \int_0^1 dw\, K(w) , \qquad (2.38)$$
$$D_1 = A_{\text{CDM}} \frac{1}{2\pi} \int_0^1 dw\, w K(w) ,$$
$$D_2 = A_{\text{CDM}} \frac{3 k_0}{32} \int_0^1 dw\, w^2 K(w) ,$$

and

$$K(w) \equiv (1-w)^2 G_Q(w) W_G(w) . \qquad (2.39)$$

Eq.(2.37) shows that, to second order in $\phi$, the QSO-galaxy angular correlation function decreases like a parabola for HDM and approximately linearly for CDM. Figs.(2a,2b)

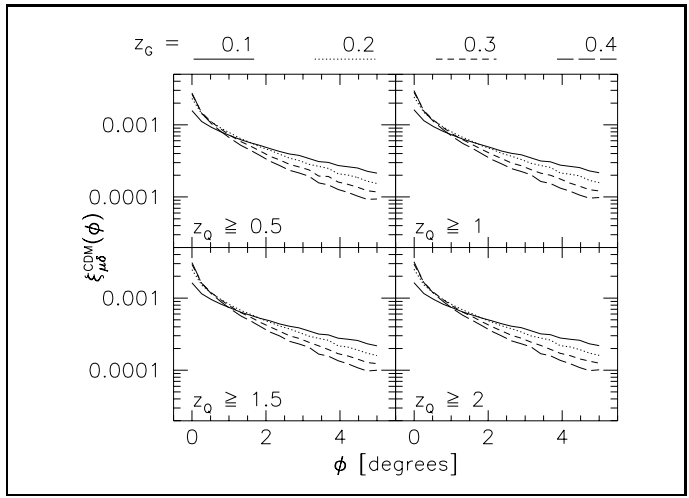

**Fig. 2a.** The angular cross-correlation function $\xi_{\mu\delta}(\phi)$ for a CDM spectrum. The four frames of the figure show results for different choices of $z_Q$ as indicated in the frames, and the line types distinguish between different choices for $z_G$ as indicated in the top line

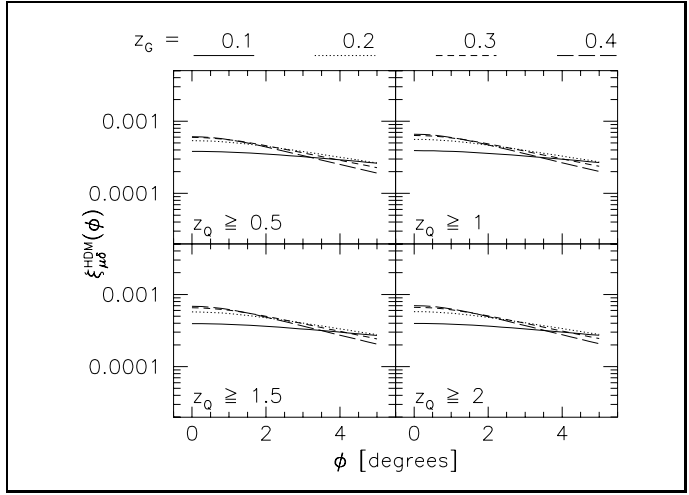

**Fig. 2b.** Similar to Fig.2a, but for an HDM spectrum

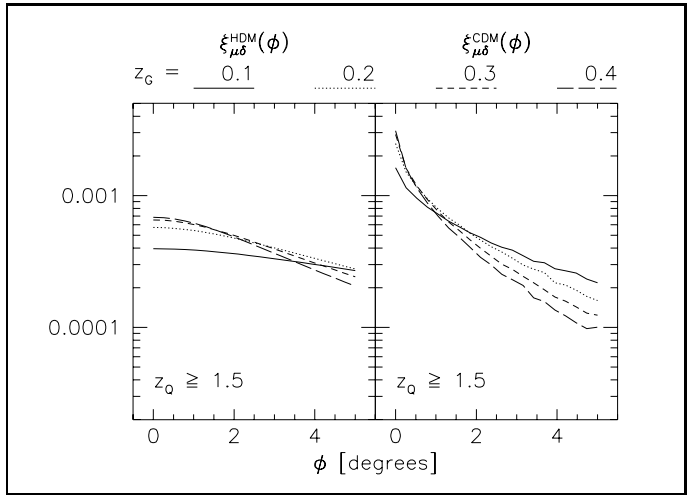

**Fig. 3.** Comparison between the angular cross-correlation function $\xi_{\mu\delta}(\phi)$ for HDM (left frame) and CDM (right frame). $z_Q$ was set to 1.5, and the line types distinguish between different choices for $z_G$, as indicated at the top of the figure. The differences in shape and amplitude for the two density-perturbation spectra are clearly seen

show $\xi_{\mu\delta}(\phi)$ for CDM and HDM for various choices of $z_G$ (indicated by the line type) and $z_Q$ (indicated in the four frames of these figures).

Figure 3 directly compares the frames for $z_Q = 1.5$ from Figs.(2a,b) to emphasize the difference in amplitude and shape of $\xi_{\mu\delta}(\phi)$ for HDM (left frame) and CDM (right frame).

Figs.(2a,2b,3) show the qualitative behaviour of $\xi_{\mu\delta}(\phi)$ derived in Eq.(2.37): The HDM correlation function starts with horizontal tangent at $\phi = 0$ and declines slowly, while the CDM correlation function starts with a rather steep decline which flattens for larger $\phi$.

## 3 Numerical simulations

This section describes numerical large-scale structure simulations performed to test whether the analytical angular cross-correlation function between QSOs and galaxies from gravitational lensing is reproduced numerically, to see how many QSOs one would need, and how deep galaxy observations should go in order to reliably estimate $\xi_{\rm QG}(\phi)$.

To simulate the large-scale structure and the galaxy distribution associated with it, I use the technique described in detail by Bartelmann & Schneider (1992). Briefly, the large-scale structure is modelled using the adhesion approximation (Gurbatov, Saichev & Shandarin 1989, Weinberg & Gunn 1990). This proves useful because this approximation renders a very fast numerical code, which allows to simulate statistically independent dark-matter distributions on a large number of lens planes. The code provides a discrete mapping from Lagrangian to Eulerian space. To find the positions of 'galaxies', the density contrast along each trajectory starting from Lagrangian space is monitored in time. When it rises above a specified threshold at some redshift for a given trajectory, this trajectory is assumed to carry a 'galaxy' later on.

Specifically, the simulation starts from CDM initial conditions normalized to $\sigma_8 = 1$ in a flat universe ($\Omega_0 = 1$, $\Lambda_0 = 0$) with $h = 1$. The simulation boxes have a comoving size of $(100~h^{-1}~{\rm Mpc})^3$. Their redshifts are chosen such that the boxes are adjacent. The matter distribution in the boxes is projected onto their midplanes, yielding the surface mass density and thus the convergence required for lensing.

The 'galaxies' are assigned luminosities drawn from the Schechter luminosity function (Schechter 1976) with the parameters determined by Efstathiou, Ellis & Peterson (1988),

$$\nu = -1.07~,~M_B^* = -19.68~,~\varphi^* = 1.56 \times 10^{-2}~{\rm Mpc}^{-3}~, \tag{3.1}$$

assuming $h = 1$. The Schechter function is cut off for luminosities below one tenth of $L^*$, and their number density is chosen such as to agree with the normalization constant $\varphi^*$. Fig.4 shows the galaxy-galaxy autocorrelation function of the resulting galaxy distribution. The figure shows that the autocorrelation function of the numerically positioned galaxies well reproduces the observed function. The magnitude-redshift relation for the model galaxies does not include $k$-corrections or luminosity evolution. For further details on the numerical large-scale structure model please refer to Bartelmann & Schneider (1992).

Light rays are propagated through the simulated stack of lens planes using the multiple lens-plane approximation (see, e.g., Schneider et al. 1992, Chap.9). Then, a synthetic QSO catalog is created assuming a random spatial distribution above redshift $z_{\rm Q} = 1$, with redshift distribution such that the fraction of QSOs observed within $dw$ of $w$ is independent of $w$ (corresponding to $\gamma = 0$ in Eq.(2.18)). Intrinsic fluxes $S'$ for these QSOs are drawn from a cumulative luminosity function of the form (2.6), with $a = -5$. QSOs are included as visible in the catalog if $\mu S'$ exceeds a specified flux threshold, where $\mu$ is the magnification factor along the line-of-sight to the QSO, and the solid-angle distortion due to lensing is taken into account by weighing the local QSO number density with $\mu^{-1}$.

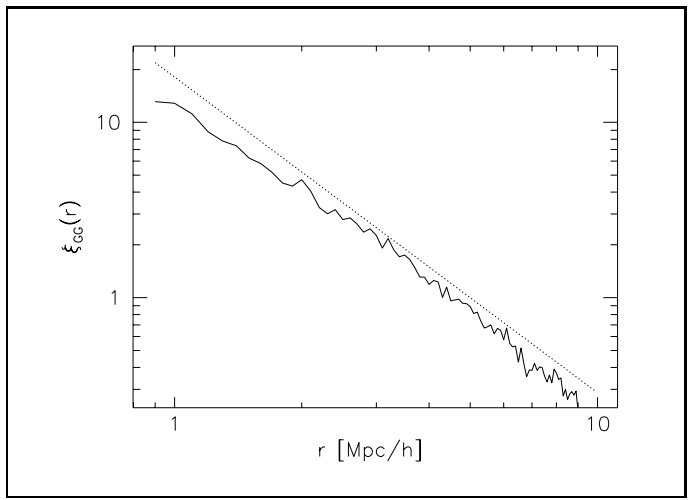

**Fig. 4.** Galaxy-galaxy autocorrelation function of the galaxy distribution created by the numerical simulation described here. The dotted line is the curve $(r/r_0)^{-1.7}$ with $r_0 = 7$ Mpc/$h$

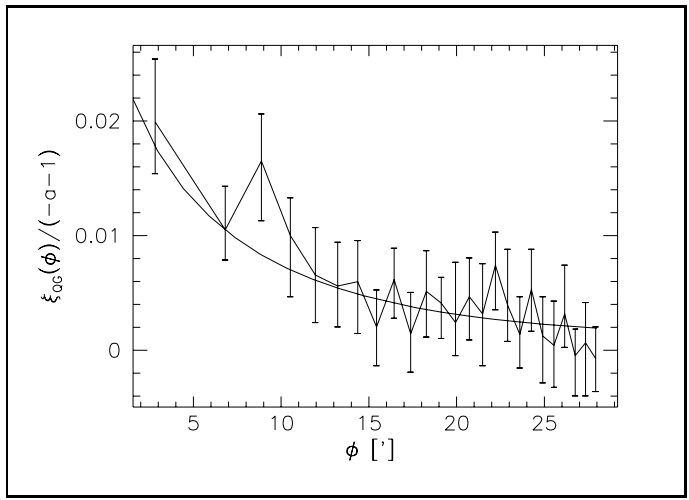

**Fig. 5.** Angular cross-correlation function between QSOs and galaxies derived from the numerical large-scale structure model described here, divided by $-a - 1 = 4$. The curve has bootstrap error bars ranging from the 10- to the 90-percentile values found for each angular separation. The smooth curve shows the analytic result for $\xi_{\mathrm{QG}}(\phi)$. Note that $h = 1$ was chosen here

From this numerically created QSO catalog, 50 subsamples of 100 QSOs each are randomly chosen. From the distribution of galaxies with magnitudes $\leq 21$ in circular fields of $28'$ radius, the angular cross-correlation function between QSOs and galaxies is determined for each of the 50 subsamples. The average number density of galaxies required for the normalization of the cross-correlation function is found by placing identical fields randomly on the simulated 'sky'. Fig.5 shows the resulting angular cross-correlation function, divided by $-a - 1 = 4$.

To compare the numerically determined cross-correlation function with the analytic result of Eq.(2.34), we have to know the bias factor of the simulated galaxy distribution. Since the galaxy-formation scheme employed here is not simple Eulerian biasing, the bias factor is not an input parameter of the simulation. Therefore, I determine $b(r)$ as the ratio of the rms fluctuations in galaxy number density and matter density (see also Eq.(3.4) below),

$$\left(\frac{\delta n_{\mathrm{G}}}{\langle n_{\mathrm{G}} \rangle}\right)\bigg|_{\mathrm{rms},r} \equiv b(r) \left(\frac{\delta \rho}{\langle \rho \rangle}\right)\bigg|_{\mathrm{rms},r} ; \tag{3.2}$$

this quantity $b(r)$ is termed 'integral spatial bias' by Cen & Ostriker (1992) and Ostriker

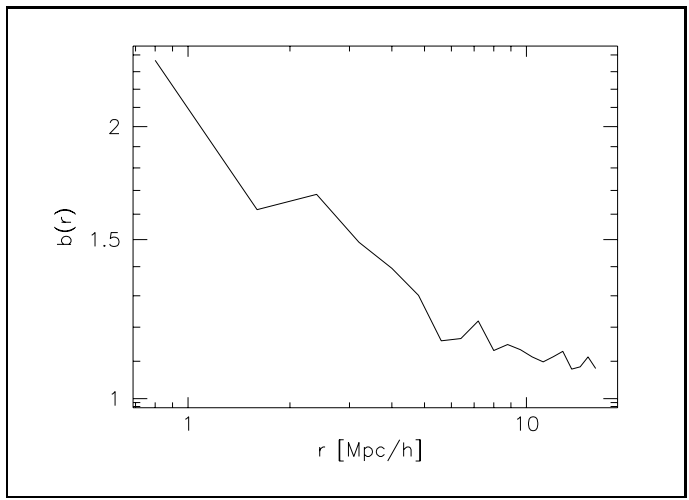

**Fig. 6.** Integral spatial bias factor $b(r)$ of the numerically simulated galaxy distribution

(1993). The so-determined $b(r)$ is displayed in Fig.6.

The figure shows that the integral spatial bias factor $b(r)$ exhibits the same qualitative behaviour as derived by Cen & Ostriker (1992), in that it approaches $\simeq 2$ at $r \simeq 1$ Mpc/$h$ and falls off towards unity for larger $r$. To incorporate a scale-dependent bias factor into the computation of the analytic angular cross-correlation function $\xi_{\mu\delta}(\phi)$ of Eq.(2.34), we employ the differential spatial bias (Cen & Ostriker 1992) defined by

$$\hat{b}(k) \equiv \sqrt{\frac{P_{\rm G}(k)}{P_\delta(k)}} \; , \tag{3.3}$$

where $P_{\rm G}(k)$ is the spectrum of the number-density fluctuations of the galaxy distribution. This quantity can be related to the integral spatial bias factor noting that the rms fluctuations in both the galaxy number density and the matter density are given by

$$\left(\frac{\delta x}{x}\right)\bigg|_{\rm rms,r} = \sigma_x(r) = \sqrt{\int \frac{d^3k}{(2\pi)^3} P_x(k) W^2(kr)} \; , \tag{3.4}$$

where $W(kr)$ is the top-hat window function

$$W(kr) = \frac{3{\rm j}_1(kr)}{kr} \tag{3.5}$$

with the spherical Bessel function of first kind ${\rm j}_1$. The symbol $\sigma_x(r)$ abbreviates the rms fluctuations of the quantity $x$ (either the galaxy number density or the matter density) on the spatial scale $r$. For an approximation, note that the window function $W(kr)$ can reasonably be approximated by the Heaviside step function ${\rm H}(1-kr)$. Then, making use of isotropy, Eq.(3.4) can be written

$$\sigma_x^2(r) \simeq \frac{1}{2\pi^2} \int_0^{1/r} dk \; k^2 P_x(k) \; . \tag{3.6}$$

Upon differentiation with respect to $r$, we find from Eq.(3.6)

$$P_x(k_r) \simeq -\frac{2\pi^2}{k_r^4} \frac{d\left[\sigma_x^2(r)\right]}{dr} \; , \tag{3.7}$$

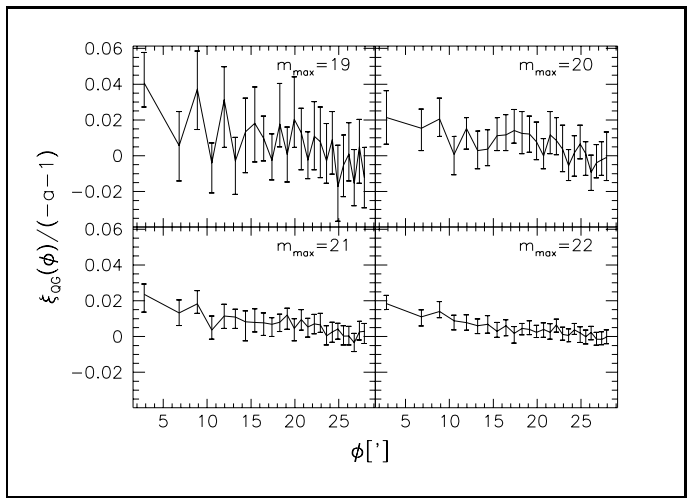

**Fig. 7.** Numerically determined QSO-galaxy cross-correlation function $\xi_{\rm QG}(\phi)$, divided by $-a-1=4$, for different depth of the synthetic galaxy 'observations' and constant QSO sample size $N_{\rm QSO}=100$. The curves in the four frames were obtained for $m_{\rm max} \in \{19, 20, 21, 22\}$ as indicated

where $k_r \equiv (1/r)$. Inserting this into Eq.(3.3) and using Eq.(3.2) yields

$$\hat{b}(k_r) \simeq b(r)\sqrt{1 - \frac{4\pi^2 r^3 \sigma_\delta^2(r)}{P_\delta(k_r)} \frac{d\ln b(r)}{d\ln r}} \;. \tag{3.8}$$

Since $b(r)$ decreases with $r$, $\hat{b}(k_r) \geq b(r)$. Having found $\hat{b}(k)$, we account for the scale-dependent bias factor by introducing the factor $\hat{b}(k)$ into Eq.(2.34), replacing $P_\delta^{(0)}(k)$ by $\hat{b}(k)P_\delta^{(0)}(k)$. Strictly speaking, $\hat{b}(k)$ can also depend on the conformal time $(1-w)$. However, since only galaxies at rather low redshifts contribute to the cross correlation between QSOs and galaxies, this possible time dependence is omitted. Finally, the dashed curve in Fig.5 was determined this way, choosing $\gamma = 0$ in the QSO weight function $G_{\rm Q}(w)$ of Eq.(2.18) to mimic the distance distribution of the synthetic QSOs in the numerical simulation, further choosing $z_{\rm Q} = 1$, and $\alpha = 2$, $\beta = 1$, $w_* = 0.3$. The latter three parameters render a good fit to the galaxy distribution of the numerical simulation. As seen in Fig.5, the analytic cross-correlation function fits the numerically determined function very well. Without accounting for the spatial dependence of the bias factor, the fit is still acceptable for $b \simeq 1.5\ldots 2$, but then the analytic cross-correlation function is slightly too shallow.

To illustrate the effects of reducing the depth of the galaxy observations and of reducing the size of the QSO sample, I show in Fig.7 the numerically determined QSO-galaxy cross-correlation function for $m_{\rm max} \in \{19, 20, 21, 22\}$, divided by $-a-1=4$, while the QSO sample size $N_{\rm QSO}=100$ is kept fixed, and analogously in Fig.8 $\xi_{\rm QG}(\phi)/(-a-1)$ for different synthtetic QSO sample sizes $N_{\rm QSO} \in \{25, 50, 75, 100\}$ and $m_{\rm max}=22$.

From the close correspondence between the analytic and the numerical cross-correlation functions in Fig.5, and from the changes in the cross-correlation function with $m_{\rm max}$ and $N_{\rm QSO}$ shown in Figs.7 and 8, I conclude that (1) the cross-correlation between background QSOs and foreground galaxies can indeed used as a tool to obtain direct information on the dark-matter density spectrum and, in particular, the bias factor, and that (2) this cross-correlation function can probably be measured observing galaxies down to $\simeq 21^{\rm st}\ldots 22^{\rm nd}$ magnitude in fields of about $20'\ldots 25'$ radius around $50\ldots 100$ QSOs. Reducing the QSO number to $\lesssim 50$, or the limiting galaxy magnitude to $m_{\rm max} \lesssim 21$, increases the error bars such that the signal can no longer be significantly

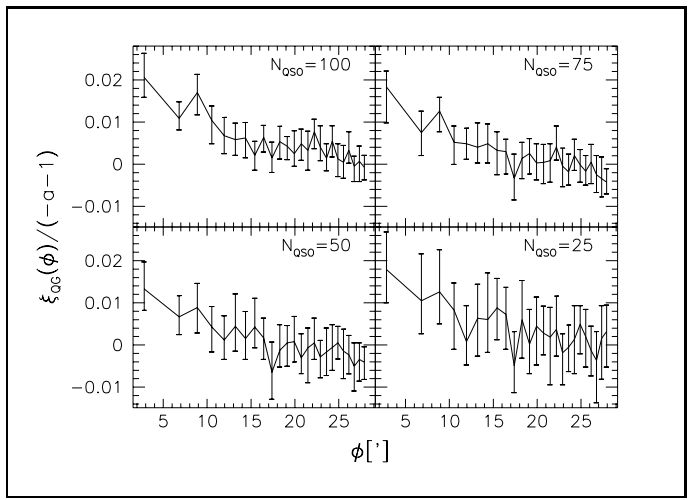

**Fig. 8.** Numerically determined QSO-galaxy cross-correlation function $\xi_{\rm QG}(\phi)$, divided by $-a-1 = 4$, for different QSO sample sizes and constant limiting galaxy magnitude $m_{\max} = 22$. The curves in the four frames were obtained for $N_{\rm QSO} \in \{25, 50, 75, 100\}$ as indicated

distinguished from zero. If the fields are reduced to much less than $\simeq 20'$ radius, the shape of the correlation function can hardly be determined, and much larger fields are hard to obtain in practice.

The size of the error bars in Figs.5, 7, and 8 are dominated by fluctuations in the galaxy number counts around the QSOs. It can be reduced by binning the galaxy counts in larger bins than used here.

Of course, the result also relies on the QSO luminosity function being sufficiently steep, since, as shown in Eq.(2.12), the QSO-galaxy cross-correlation function is proportional to $(a - 1)$. For $a = -1$, the result would be zero. However, I show in the next section that for the 1-Jansky sample of radio-loud QSOs, the double-waveband magnification bias renders effective values for $a$ which are compatible with the value $a = -5$ chosen here.

## 4 Double-waveband magnification bias

It was first argued by Borgeest et al. (1991) that the magnification bias is more efficient in a sample of sources which is flux-limited in two or more wavebands rather than in only one, provided the fluxes in these wavebands are mutually uncorrelated. The reason is easy to see. Consider a sample of sources which is flux limited in $n$ different wavebands. The unlensed counts of these sources depend on $n$ fluxes $S'_i$. Assuming that the magnification factor in all $n$ wavebands is the same, Eq.(2.7) reads

$$n_{\rm Q}(S_1, \ldots, S_n) = \frac{1}{\mu}\, n'_{\rm Q}\left(\frac{S_1}{\mu}, \ldots, \frac{S_n}{\mu}\right)\;. \tag{4.1}$$

The latter assumption that the magnification factor is the same in all $n$ wavebands is not trivial despite the achromaticity of lensing because the radiation in different wavebands can be emitted from regions of different size of the sources. Microlensing, e.g., can only affect small parts of the source like the continuum-emitting region, while the radio emission usually comes from much larger regions whose flux remains unaffected.

The double-logarithmic slope of the number counts effective for the magnification bias is given by

$$1 + a_{\mathrm{eff}} = -\frac{d \ln n_{\mathrm{Q}}(S_1, \ldots, S_n)}{d \ln \mu} = 1 + \sum_{i=1}^{n} \frac{\partial \ln n'(S'_1, \ldots, S'_n)}{\partial \ln S'_i} \, , \qquad (4.2)$$

where Eq.(4.1) was used. As argued above, the magnification factors induced by the lensing effect of large-scale structures is close to unity, $\mu = 1 + \delta\mu$ with $\delta\mu \ll 1$. Therefore, only the local behaviour of the unlensed number counts is relevant for the present consideration. If the number counts can locally be approximated by power laws with index $c_i$,

$$\frac{\partial \ln n'_{\mathrm{Q}}(S'_1, \ldots, S'_n)}{\partial \ln S'_i} = c_i \, , \qquad (4.3)$$

and if the unlensed fluxes $S'_i$ are mutually independent, it follows from Eq.(4.2) that the effective double-logarithmic slope is just the sum of the slopes in the $n$ wavebands,

$$a_{\mathrm{eff}} = \sum_{i=1}^{n} c_i \, . \qquad (4.4)$$

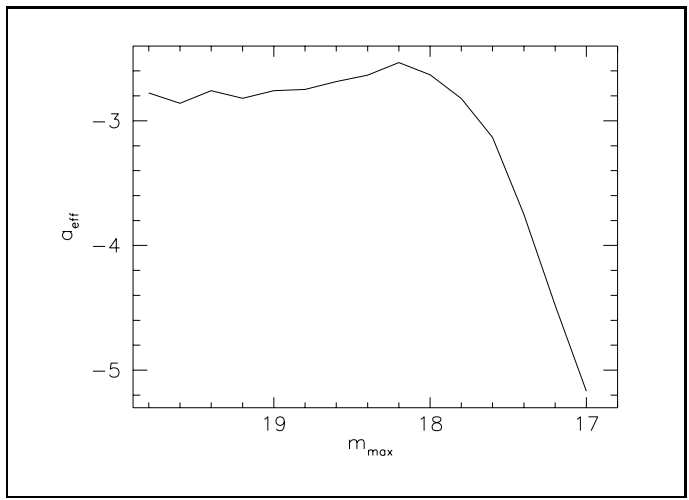

**Fig. 9.** Effective cumulative slope of 1-Jansky sources with redshift $z \geq 1$ as a function of the limiting optical magnitude $m_{\max}$. For optically bright subsamples, the effective number-count slope decreases to $a_{\mathrm{eff}} = -4 \ldots -5$

In any case, Eq.(4.2) shows that imposing flux thresholds in two or more wavebands effectively steepens the number-count distribution.

This steepening of the effective number counts is demonstrated in Fig.9 using the optically identified subsample of the 1-Jansky sample of radio-loud QSOs with redshift $z \geq 1$. These are by definition flux-limited to $S_{\mathrm{rad}} \geq 1$ Jy at 5 GHz, and since they are optically identified, one can additionally impose an optical flux limit. For the figure, optical magnitudes $m$ of the 1-Jansky sources are converted to fluxes relative to some arbitrary reference magnitude $m_{\max}$,

$$S_{\mathrm{opt}} = 10^{0.4(m - m_{\max})} \, . \qquad (4.4)$$

For $S \geq 1$, the cumulative slope $a_{\mathrm{eff}}$ of the counts $P_{\mathrm{opt,rad}}(S)$ of 1-Jansky sources with

$$\min(S_{\mathrm{opt}}, S_{\mathrm{rad}}) \geq S \qquad (4.5)$$

is determined and plotted as a function of $m_{\max}$. The number-count slope without optical flux limit is $a_{\mathrm{rad}} = -2.3$.

The figure shows that subsamples of the 1-Jansky sources can have a steep effective number-count slope when they are flux-limited in the optical wavelength in addition to the intrinsic flux limitation in the radio waveband at 5 GHz.

# 5 Summary and discussion

It was shown in previous papers that there is evidence for correlations on angular scales of several 10 arc minutes between background QSOs and foreground galaxies on a high significance level (Fugmann 1992, Bartelmann & Schneider 1993b, 1994a, Bartelmann et al. 1994b). As argued there and in Bartelmann & Schneider (1993a), gravitational lensing by large-scale structures provides a viable explanation for these correlations. If this interpretation is valid, it proves the biasing hypothesis of galaxy formation and provides a method to trace extended dark structures by their gravitational lens effect, which would otherwise be hard to detect.

This paper develops this idea further. I have computed the angular two-point cross-correlation function $\xi_{\mathrm{QG}}$ between background sources and foreground galaxies due to the gravitational lensing effect of large-scale structures. This correlation function is sensitive to the power spectrum of dark-matter fluctuations and therefore potentially allows to discriminate between different cosmogonic models. As shown in Eqs.(2.12), (2.37), and (2.38), $\xi_{\mathrm{QG}}$ is proportional to the spectral amplitude $A$, to the bias factor $b$ of galaxy formation, and to $(-a-1)$, where $a$ is the double-logarithmic slope of the cumulative QSO number counts. The spectral amplitude $A$ is known from the analysis of the CMB fluctuations, and the slope $a$ can independently be determined from the background QSO sample employed. Thus, $\xi_{\mathrm{QG}}$ also allows to measure the bias factor $b$.

Numerical simulations suggest that $\xi_{\mathrm{QG}}$ can be measured using a sample of $50\ldots100$ QSOs with redshift $z \geq 1$ if galaxy observations down to a limiting magnitude of $21\ldots22$ are available in fields of about $25'$ radius around these QSOs. The slope of the QSO number counts can effectively be increased by imposing flux limits in two or more wavebands upon the QSO sample. This multiple-waveband magnification bias (Borgeest et al. 1991) can increase the effective number-count slope to $a_{\mathrm{eff}} \gtrsim -5$ in the 1-Jansky sample of high-redshift, radio-loud QSOs.

There are several problems which may render the determination of $\xi_{\mathrm{QG}}$ difficult in practice. First, $\xi_{\mathrm{QG}}$ has to be properly normalized. To do so, the average galaxy number density has to be determined, a measurement prone to observational errors. However, since $\xi_{\mathrm{QG}}$ is an *angular* correlation function, it is much easier to measure than the *spatial* galaxy autocorrelation function, because errors in the distance determination of galaxies do not affect the result. Even if the amplitude of $\xi_{\mathrm{QG}}$ is not determined correctly, its shape can still yield information about the dark-matter perturbation spectrum. Second, galaxy counts and optical QSO fluxes are influenced by dust absorption. Although the 1-Jansky QSO sample is radio selected, this effect could become important when one imposes an additional optical flux limit to take advantage of the multiple-waveband magnification bias. More severe, however, is the influence of spatially varying dust absorption or reddening on the galaxy number counts. This could affect the normalization and the shape of $\xi_{\mathrm{QG}}$. In order to minimize such effects, the galaxy counts required for determining $\xi_{\mathrm{QG}}$ should be done in red light or near infrared. Third, the correlation scale investigated here requires large CCD frames or CCD mosaics. While sufficiently large CCDs will probably become available in the near future, CCD mosaics pose the problem of adapting galaxy counts to a common level in all their parts.

Despite these observational problems, the measurement proposed here provides a tool to determine the shape of the power spectrum of density fluctuations independent of other attempts, and to quantify the biasing hypothesis of galaxy formation; the latter being hard to fulfil in a different way.

*Acknowledgements.* I wish to thank Nick Kaiser and Peter Schneider for fruitful discussions and valuable comments.

# Appendix A. The deflection angle of a weakly perturbed light ray

We restrict ourselves to a model universe which is on average of the Einstein-de Sitter type ($\Omega_0 = 1$, $\Lambda_0 = 0$, $p = 0$) and whose density is weakly perturbed by the relative density contrast $\delta$ with $\langle\delta\rangle = 0$ when averaged over sufficiently large volumes. In close analogy to Kaiser (1992), we write the space-time metric in the form

$$ds^2 = a^2(\eta)\left[d\eta^2 - \delta_{ij}dw^i dw^j\right] , \tag{A1}$$

where $\eta$ is the conformal time which is defined by

$$d\eta = \frac{c\,dt}{a(t)} = \frac{c\,da}{a\dot{a}} \tag{A2}$$

and the condition that $\eta = 0$ when $t = 0$. Friedmann's equation reads, for our choice of the cosmological parameters,

$$\dot{a} = \frac{H_0}{\sqrt{a}} , \tag{A3}$$

and since $a(z) = (1+z)^{-1}$ when $a$ is normalized to unity at present, $a_0 \equiv a(t_0) = 1$, we find from Eqs.(A2) and (A3)

$$\eta(z) = \frac{2c}{H_0}\frac{1}{\sqrt{1+z}} . \tag{A4}$$

Thus, the comoving time has the dimension of a length. If we express in the following all lengths in units of twice the Hubble length, we can write

$$\eta(z) = \frac{1}{\sqrt{1+z}} = \sqrt{a(z)} . \tag{A5}$$

The comoving distance $w$ along a radial light ray starting at the observer position follows from $ds = 0$, or, from Eq.(A1),

$$d\eta = -dw , \tag{A6}$$

and therefore the comoving distance equals the conformal lookback time,

$$1 - \eta(z) = w(z) = 1 - \frac{1}{\sqrt{1+z}} . \tag{A7}$$

The density contrast $\delta$ will give rise to gravitational lens effects on light rays. It satisfies Poisson's equation, which reads, in comoving coordinates,

$$\Delta\Phi = 4\pi G\bar{\rho}a^2\delta ; \tag{A8}$$

see, e.g., Peebles [1993, Eq.(5.107)]. $\bar{\rho}$ is the mean density of the background universe, hence

$$\bar{\rho} = a^{-3}\bar{\rho}_0 = \frac{3H_0^2}{8\pi G a^3} \ , \qquad (A9)$$

and thus Eq.(A8) transforms to

$$\Delta\Phi = \frac{3H_0^2}{2a}\delta \ . \qquad (A10)$$

Measuring lengths in units of twice the Hubble length and the potential in units of $c^2$, we can write instead of Eq.(A10)

$$\Delta\Phi(\vec{w},\eta) = \frac{6}{a}\delta(\vec{w},\eta) = \frac{6}{\eta^2}\delta(\vec{w},\eta) \ , \qquad (A11)$$

where we have inserted the arguments for clarity: $\vec{w}$ is the comoving position considered, and $\eta$ is the conformal time.

Consider now a radial light ray starting at the observer position into direction $(\theta_1, \theta_2)$; then, $\vec{w} = (w\theta_1, w\theta_2, w)$ and $\eta = 1 - w$. This light ray will be deflected by the gravitational-lens effect of the density perturbations according to the equation of motion

$$\ddot{r}_j(\vec{w},\eta) = -2\Phi_{,j}(\vec{w}, 1-w) \ , \qquad (A12)$$

where $\vec{r}$ is the comoving deviation vector of the light ray relative to the unperturbed ray, and the comma preceding the index $j$ denotes the partial derivative with respect to $r_j$.

The total deviation from the unperturbed ray is then given by

$$r_j(\theta_1,\theta_2,w) = -2\int_0^w dw' \int_0^{w'} dw'' \Phi_{,j}(\vec{w}'', 1-w'') \ . \qquad (A13)$$

This inner integral is easily evaluated along the *unperturbed* light ray; it yields

$$r_j(\theta_1,\theta_2,w) = -2\int_0^w dw'(w-w')\Phi_{,j}(w'\theta_1, w'\theta_2, w'; 1-w') \ . \qquad (A14)$$

The deflection angle of the light ray propagating into the direction $\vec{\theta} = (\theta_1, \theta_2)$ is given by $w\alpha_j = r_j$, or

$$\alpha_j(\vec{\theta},w) = -\frac{2}{w}\int_0^w dw'(w-w')\Phi_{,j}(w'\theta_1, w'\theta_2, w'; 1-w') \ . \qquad (A15)$$

This is Eq.(2.1) for the deflection angle of a weakly perturbed light ray propagating through an Einstein-de Sitter background universe.